\documentclass{optica-article}

\journal{opticajournal} 

\articletype{Research Article}

\usepackage{lineno}
\usepackage{comment}
\usepackage[normalem]{ulem}

\usepackage{array}

\begin{document}

\title{Enabling data-driven and bidirectional model development in Verilog-A for photonic devices}

\author{Dias Azhigulov\authormark{1,*}, Zeqin Lu\authormark{2}, James Pond\authormark{2}, Lukas Chrostowski\authormark{1}, and Sudip Shekhar\authormark{1}}

\address{\authormark{1}Department of Electrical and Computer Engineering, University of British Columbia, 2332 Main Mall, Vancouver, BC V6T 1Z4, Canada\\
}
\address{\authormark{2}Ansys Lumerical Inc., 1095 W Pender St \#1700, Vancouver, BC V6E 2M6, Canada}
\email{\authormark{*}dias.azhigulov@ece.ubc.ca} 


\begin{abstract*} 
We present a method to model photonic components in Verilog-A by introducing bidirectional signaling through a single port. To achieve this, the concept of power waves and scattering parameters from electromagnetism are employed. As a consequence, one can simultaneously transmit forward and backward propagating waves on a single wire while also capturing realistic, measurement-backed response of photonic components in Verilog-A. We demonstrate examples to show the efficacy of the proposed technique in accounting for critical effects in photonic integrated circuits such as Fabry-Perot cavity resonance, reflections to lasers, reflection cancellation circuits, etc. Our solution makes electronic-photonic co-simulation more intuitive and accurate.
\end{abstract*}

\section{Introduction}

Photonic Integrated Circuits (PICs) play a critical role in many modern applications, ranging from communication to sensing \cite{shekhar2023silicon}. The rise of CMOS compatible silicon photonic processes accelerated this trend. Practically, in each case, they are accompanied by electronics, either monolithically integrated into a single chip or otherwise. This necessitates a common simulation platform where one can simulate PICs and Electronic Integrated Circuits (EICs) to capture their interactions.

Historically, photonic and electronic circuit simulation engines (e.g. Lumerical INTERCONNECT \cite{INTC} and Cadence Spectre \cite{Virtuoso}) evolved independently, each limiting the simulation to their respective domains. The electronic-photonic co-simulation aims to bridge the gap between the two domains through various tools. It can be broadly categorized into two groups: Electronic Photonic Design Automation (EPDA) simulation and Electronic Design Automation (EDA) simulation. An example of the former is the Cadence-Lumerical EPDA platform~\cite{CadLum} that conjoins two different simulators by exchanging simulation data back and forth between the two. It provides an accurate and versatile co-simulation platform, but transient simulations for circuits incorporating feedback can be slow. The EDA-only co-simulation leverages hardware description languages (HDLs) such as Verilog-A for compact modeling of photonic devices to support PICs natively in SPICE-class electrical circuit simulators~\cite{kononov, Cheryl, jleu, Fang:24, 6842123, bwang, DEFOUCAULD2023108538}. Since the number of electronic devices far exceeds their photonic counterparts in any silicon photonic application~\cite{shekhar2023silicon}, Verilog-A-based native simulations bring ease and simplicity to system architects and CMOS designers. Kononov et al.~\cite{kononov} laid the groundwork for photonic component modeling, capturing effects such as phase shift, delay, attenuation, and electrical bandwidth. The follow-up works \cite{Cheryl, jleu} further improved the models and introduced new features such as multi-wavelength communication using a single wire and laser phase noise. They also presented complex system-level EIC and PIC co-simulation setups. 

Silicon photonics integrated circuits operate at a carrier frequency of 100s of THz, and therefore must be treated as transmission lines. Therefore, capturing the impact of reflections is crucial for various circuits. For example, back-reflections from a PIC into a laser without an isolator can degrade the laser performance, and even destabilize it~\cite{Petermann88,ShomanJLT}. Circuit techniques to stabilize the lasers~\cite{Doerr, Dong15, Hauck, ShomanJLT} or improve its linewidth using self-injection locking~\cite{Kondratiev} must capture reflections. Cavity-based optical filters~\cite{Patel_Ghillino_Korthorst_2021} must account for reflection for accurate modeling of the filter frequency response. 

The modeling of reflections and bidirectional signaling in Verilog-A has remained incomplete, even leading to claims in~\cite{Patel_Ghillino_Korthorst_2021} that Verilog-A-based co-simulation is not meant to handle such features. Nevertheless, several attempts have been made to address bidirectional signal propagation. 
Several models~\cite{Cheryl, jleu, Fang:24, 6842123, bwang} handle the forward and backward propagating waves in separate wires. However, they do not account for reflections from all ports (e.g. in directional couplers). Additionally, reflection-centered devices such as Bragg gratings are not trivial to implement in Verilog-A. The previous methods are limited to what they can model since they all base it on analytical equations. This leads to simplifications in modeling and idealistic simulations.

Recently, de Foucauld et al.~\cite{DEFOUCAULD2023108538} presented a technique to handle reflections in the same wire as the transmitted signal without the two corrupting each other. The presented test bench is, however, very simple and not representative of more intricate photonic circuits. The authors state that their method does not hold up for the cases where multiple optical sources are present.  

We present a methodology to capture forward and backward propagating signals using the same wire, without putting any limitations on all prior modeling techniques~\cite{LumPatent}. Through this work, one only needs to use half as many wires as being used in traditional Verilog-A photonic models. Moreover, by taking advantage of the bidirectional signaling technique we build measurement data-driven models for passive photonic devices using scattering parameters (\textit{s-parameters}) \cite{pozar2011microwave}. Unlike the previous methods~\cite{Cheryl, jleu, Fang:24, 6842123, bwang}, this lets users recreate response of any arbitrary passive photonic element in Verilog-A without the need for device physics expertise that is required for conventional approach. This technique naturally captures fabrication induced imperfections as well. While both the bidirectionality and data-driven modeling features have been present in electrical circuit simulators (e.g. transmission line and BSIM transistor models), their application has not been shown for user-defined custom photonic components such as the ones based on Verilog-A, which limits the modeling potential of the HDL. Therefore, with this work, we demonstrate how to integrate the aforementioned features into Verilog-A, thereby enabling the modeling of a wider range of photonic devices and systems. As a final note, we highlight all the differences between this work and the prior arts in Table \ref{tab:my-table}.

The paper consists of four sections. Section II describes the methodology of bidirectional signaling in Verilog-A with a brief theoretical background. In section III we simulate and present the results of circuits consisting of Verilog-A components to prove the viability of our solution. Section IV concludes our work.

\begin{table}[]
\begin{tiny}
\caption{Comparison between this work's features and prior arts.}
\resizebox{\textwidth}{!}{%
\begin{tabular}{|
>{\columncolor[HTML]{C0C0C0}}c |c|c|c|c|}
\hline
 & \cellcolor[HTML]{C0C0C0}\textbf{[6]} & \cellcolor[HTML]{C0C0C0}\textbf{[7]} & \cellcolor[HTML]{C0C0C0}\textbf{[9]} & \cellcolor[HTML]{C0C0C0}\textbf{This work} \\ \hline
\textbf{Model types} & \cellcolor[HTML]{FFD700}\begin{tabular}[c]{@{}c@{}}\vspace{-4pt}Analytical \\ and curve-fit \end{tabular} & \cellcolor[HTML]{FFD700}Analytical & \cellcolor[HTML]{FFD700}\begin{tabular}[c]{@{}c@{}}\vspace{-4pt}Analytical \\ and curve-fit \end{tabular} & \cellcolor[HTML]{32CB00}\begin{tabular}[c]{@{}c@{}}\vspace{-4pt}Analytical, curve-fit, and \\ measurement data based\end{tabular} \\ \hline
\cellcolor[HTML]{C0C0C0}\textbf{Bidirectional signaling} & \cellcolor[HTML]{F56B00}2 ports & \cellcolor[HTML]{F56B00}2 ports & \cellcolor[HTML]{32CB00}1 port & \cellcolor[HTML]{32CB00}1 port \\ \hline
\textbf{Reflections} & \cellcolor[HTML]{F56B00}No & \cellcolor[HTML]{FFD700}Yes & \cellcolor[HTML]{FFD700}Yes & \cellcolor[HTML]{32CB00}Yes, from all ports \\ \hline
\textbf{Fabrication effects} & \cellcolor[HTML]{F56B00}No & \cellcolor[HTML]{F56B00}No & \cellcolor[HTML]{F56B00}No & \cellcolor[HTML]{32CB00}Yes \\ \hline
\textbf{\begin{tabular}[c]{@{}c@{}}\vspace{-4pt}Supports multiple \\ optical sources\end{tabular}} & \cellcolor[HTML]{32CB00}Yes & \cellcolor[HTML]{32CB00}Yes & \cellcolor[HTML]{F56B00}No & \cellcolor[HTML]{32CB00}Yes \\ \hline
\end{tabular}%
}
\label{tab:my-table}
\end{tiny}
\end{table}

\section{Methodology}

\subsection{Theory}
Before delving into the details of our implementation, it should be emphasized that at the baseline one can use the established techniques \cite{kononov,jleu} of photonic component modeling in Verilog-A. Our bidirectional signaling methodology simply encapsulates the behavioural description code of a device.

There is a large body of literature on analytical description of forward and backward propagating waves in electromagnetics theory \cite{pozar2011microwave}. One very pertinent concept to our problem is \textit{power waves} and they are defined as:

\begin{equation}
    a = \frac{V+Z_RI}{2\sqrt{R_R}}
\label{eq:1}
\end{equation}
\begin{equation}
    b = \frac{V-Z_R^*I}{2\sqrt{R_R}},
    \label{eq:2}
\end{equation}
where $a$ and $b$ are forward and backward propagating wave amplitudes, respectively, V and I are the voltage and current present in a circuit. $Z_R = R_R + jX_R$ is the reference impedance with resistance of $R_R$ and reactance of $X_R$. For simplicity, we can assume that our reference impedance is purely real, $Z_R = R_R$, and its value can be arbitrary. Solving for V and I in Eq. \ref{eq:1} and \ref{eq:2} we get:

\begin{equation}
    V = \frac{Z_R^* a+Z_R b}{\sqrt{R_R}}=\sqrt{R_R}(a+b)
\label{eq:3}
\end{equation}
\begin{equation}
    I = \frac{a-b}{\sqrt{R_R}}
    \label{eq:4}
\end{equation}

\subsection{Implementation}

\begin{figure}[h!]
\centering\includegraphics[width=11cm]{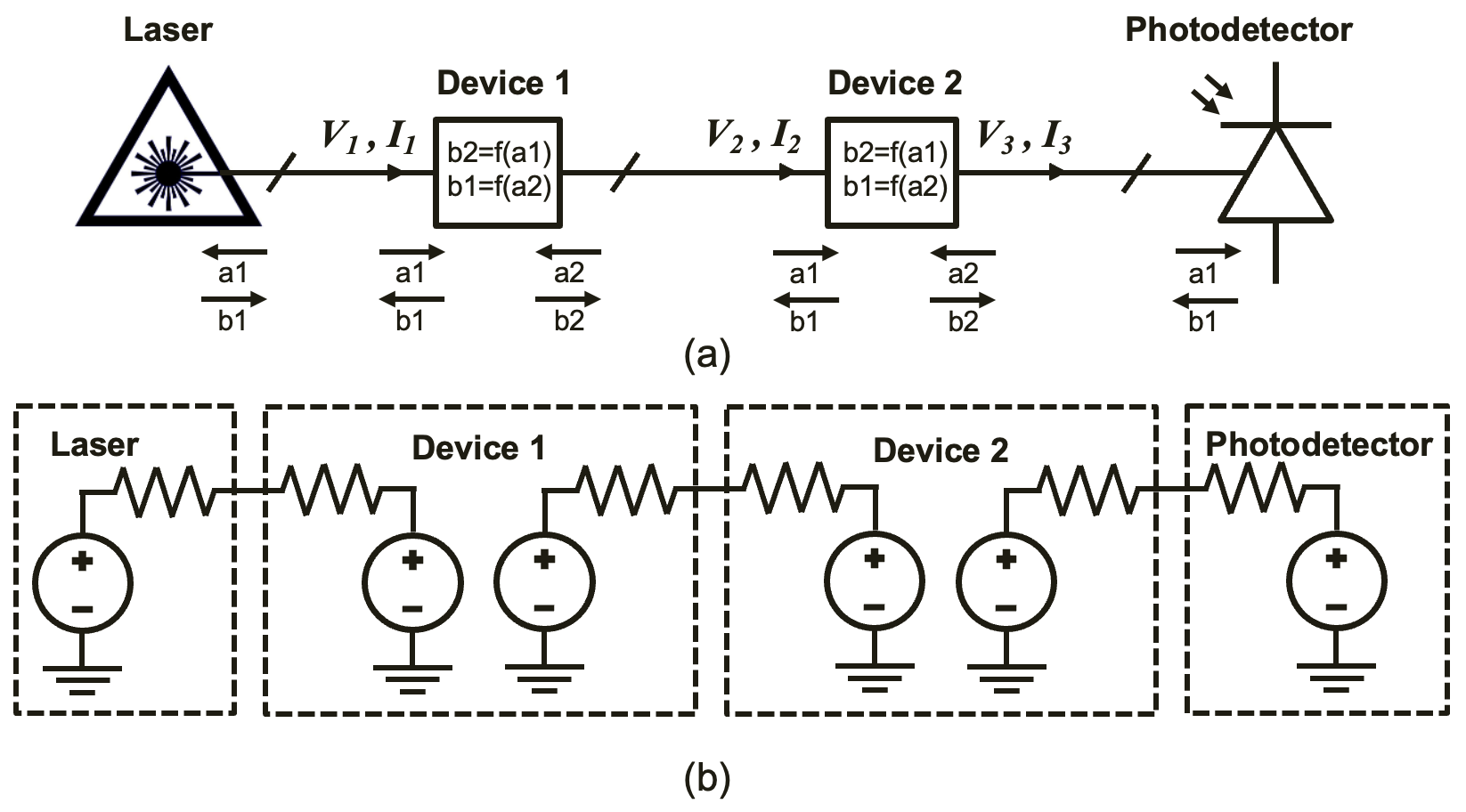}
\caption{(a) Illustration of a photonic circuit; (b) Its electronic equivalent in Verilog-A.}
\label{fig:2}
\end{figure}

Next, we explain how to incorporate the power waves into Verilog-A. Consider the basic example of a photonic circuit and its electrical counterpart shown in Fig. \ref{fig:2}a and \ref{fig:2}b. The former consists of the laser, photodetector, and a couple of black box elements, which may perform arbitrary functions on the input signals. In the latter, however, we describe every circuit element using its interfaces, that is, ports. Each port is a voltage source in series with a resistance (or more generally, an impedance). The sources can supply arbitrary voltages depending on the component behavior. In combination with the resistors, they allow current and voltage flow in the circuit thereby making bidirectional signaling possible. While various resistance values can be chosen to mitigate potential convergence issues, the key here is to set all of them to the same value $R_R$.

To implement the bidirectionality in Verilog-A, one has to use the bidirectional ports (\textit{inout} construct) in Verilog-A when defining all the optical ports. Afterwards, for any component, one can calculate the incoming wave amplitude at a given port using the voltage and current readings at that port and plugging them into Eq. \ref{eq:1}. Then, the outgoing wave can be set as a function of the incoming wave ($b = f(a)$). There are no restrictions, besides the ones imposed by Verilog-A itself, on what kind of function this is. Following this, we compute either the new voltage or current values at each port through Eq. \ref{eq:3}-\ref{eq:4}.  In the context of the photonic circuit depicted in Fig. \ref{fig:2}a, it can be noticed that there are forward and backward propagating waves at each port of a given device. Additionally, the voltage and current values vary from one node to another. Thus, while voltages and currents carry the information, the wave values $a$ and $b$ are the ones that represent the actual optical field. Such freedom in functionality and independence from  node to node can occasionally lead to convergence issues due to widely varying currents and voltages throughout the circuit. If that is the case, controlling the signal's tolerance levels in Verilog-A can help alleviate the problem \cite{verilogAMS}.

It should also be highlighted that for transmission lines, reflection takes place whenever there is a mismatch in impedance between two mediums (e.g. characteristic impedance of transmission line $\neq$ load impedance). For power waves, reflection occurs when the reference impedance is not the same as the load impedance~\cite{pozar2011microwave}:

\begin{equation}
    \Gamma_p = \frac{b}{a} = \frac{V-Z_R^* I}{V+Z_R I} =  \frac{Z_L-Z_R^*}{Z_L+Z_R},
    \label{eq:5}
\end{equation}
where $\Gamma_p$ is the reflection coefficient of the power waves. In our Verilog-A models, we set $Z_L$ and $Z_R$ to the same real value. In this way, there are no reflections by default. The only source of reflections are the intermediate black box components. 

To incorporate reflections into the black box models in Fig.~\ref{fig:2}, one can use analytical descriptions of the plane wave transmission and reflection off an interface \cite{DEFOUCAULD2023108538} or one can utilize scattering matrices as an extension of the power waves concept \cite{pozar2011microwave}. The former has already been done in Verilog-A in~\cite{DEFOUCAULD2023108538}. We will only explore the latter approach due to its additional merits. Scattering matrix consists of s-parameters, which describe the response of linear electrical or optical networks at the port $i$ to a certain input signal at the port $j$ \cite{säckinger2005broadband}. For a two port network/device it can be expressed as:

\begin{equation}
    S = \begin{bmatrix}
s_{11} & s_{12}\\
s_{21} & s_{22} 
\end{bmatrix}  \quad and \quad     s_{ij} = \frac{b_i}{a_j},
\label{eq:6}
\end{equation}
where $b_i$ is the outgoing wave at the port $i$ and $a_j$ is the incoming wave at the port $j$. The introduction of s-parameters is also advantageous for data-driven model development since all the photonic blocks can be characterized by the frequency dependent s-parameter data. We note that the impedance at which these devices are measured is not necessarily related to the reference impedance $R_R$. The latter is used as a knob to control the reflectivity of a component from each port in the absence of measurement data. In Verilog-A, one can import s-parameters as a Lookup Table (LUT) and write the equations for the outgoing waves using the s-parameters \textcolor{red}{\cite{veriaref}}. For a generic \textit{n}-port network this can mathematically be described as:

\begin{equation}
\begin{bmatrix}
    b_1 \\
    \vdots \\
    b_n \\
\end{bmatrix} = 
\begin{bmatrix}
s_{11} & \cdots & s_{1n}\\
\vdots &  \ddots & \vdots \\
s_{n1} & \cdots & s_{nn} 
\end{bmatrix}
\begin{bmatrix}
    a_1 \\
    \vdots \\
    a_n \\
\end{bmatrix}
\end{equation}

Thus, the outgoing waves for the two port devices from Fig. \ref{fig:2} are expressed as $b_1 = s_{11} a_1+s_{12} a_2$ and $b_2 = s_{21} a_1 + s_{22} a_2$. From the definition of s-parameters, it can be deduced that $s_{11}$ is the reflection coefficient at port 1 and $s_{21}$ is the transmission coefficient. Consequently, $b_1$ now captures the transmitted signal from the opposite end as well as the reflected signal from the same port, giving rise to forward and backward propagating waves in our circuit. It should be emphasized that not every photonic device needs the s-parameter based model description (e.g. a homogeneous medium such as ideal waveguides does not produce any reflections). Using Eq. \ref{eq:1}-\ref{eq:4} is enough to run bidirectional simulations in any commercial EDA software. On the other hand, if the user wants to preserve the parametric nature of photonic components, they can leverage the analytical modeling approach. Another solution is to enable interpolation between multiple s-parameters.

We note that all the commercial EDA tools already handle reflections when simulating electronic devices with transmission lines. The only caveat is that those tools keep everything in voltage and current domains. Thus, in order to observe reflections one has to interpret the voltage and current waveforms, which come as a superposition of transmitted and reflected signals. Since our situation is similar, we implement a monitoring device to read the optical power and phase of forward and backward propagating signals in a given node. A similar device can also be found in photonic circuit simulators such as Lumerical INTERCONNECT \cite{INTC}.

\section{Simulation}

\subsection{Michelson Interferometer testbench}
\begin{figure}[h!]
\centering\includegraphics[width=13cm]{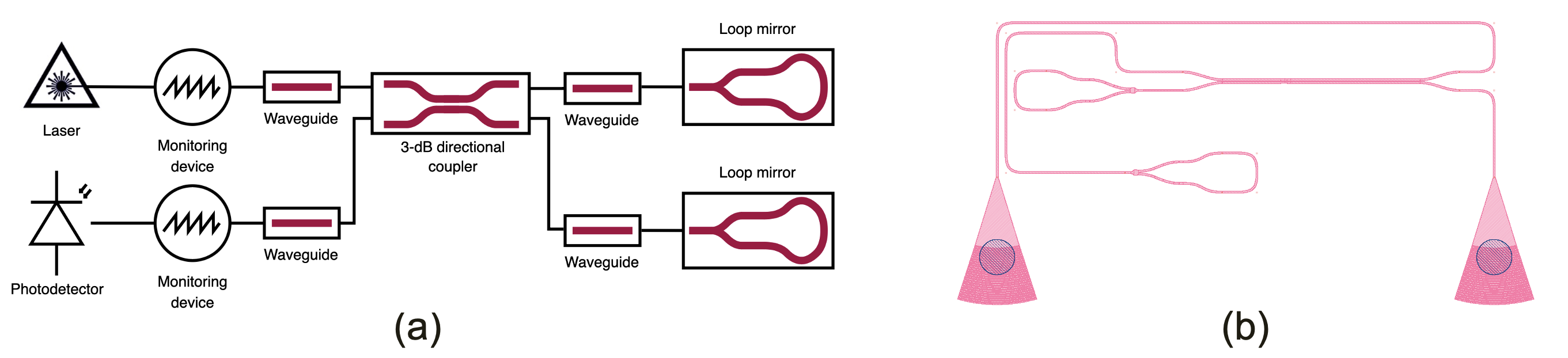}
\caption{(a) Schematic diagram of the MI built using Verilog-A components; (b) Layout of the measured MI.}
\label{fig:2b}
\end{figure}

\begin{figure}[h!]
\centering\includegraphics[width=13cm]{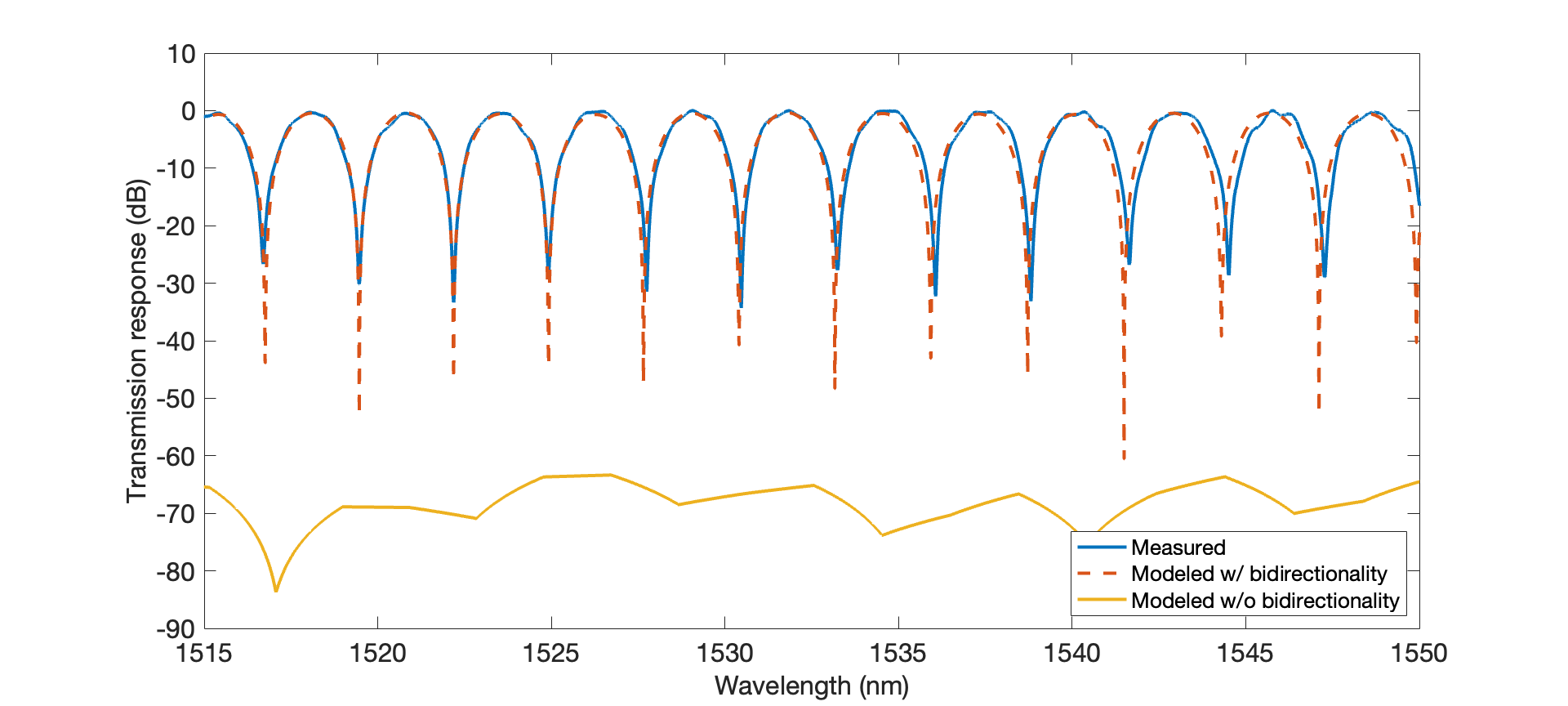}
\caption{Transmission response of the MI.}
\label{fig:4}
\end{figure}

To demonstrate the efficacy of our modeling approach, we built a Michelson Interferometer (MI) using a Verilog-A library of silicon photonic devices \cite{kononov,jleu}. As per the typical MI design, it consists of 3-dB directional coupler, waveguides, and loop mirrors built from Y-branch and waveguide bends. Note that the laser is internally set to have an isolator in order to keep feeding constant power, though this can be modified in the code. In our library, all the passive photonic device models except the waveguide employ the data-driven approach and are built using the data from \cite{ebeam}. The waveguide is described using analytical equations.  Active components such as photodetectors and modulators are developed through a combination of analytical description, curve-fits, and simulation/measurement data \cite{kononov, chrostowski_hochberg_2015, PD}. Thermal phase shifters are based on the analytical equations presented in \cite{jleu}. Laser is the only device in our library that has purely behavioural description because we use it as a continuous wave (CW) supply to the PICs \cite{kononov}. Fig. \ref{fig:2b}a depicts the schematic view of the MI assembled in Cadence Virtuoso \cite{Virtuoso}. The corresponding layout of the fabricated device is shown in Fig. \ref{fig:2b}b. MI is an ideal testbench for demonstration of bidirectional signaling since the forward and backward propagating signals are simultaneously present in each arm. Furthermore, signals in both arm should constructively interfere to be able to see the signal at the photodetector side. In case of destructive interference, all of the returning signals should appear at the laser side. In contrast to the example from~\cite{DEFOUCAULD2023108538}, our testbench can be viewed to have three optical sources (one forward propagating signal from the laser and two reflected signals from the mirror loops) all meeting at the directional coupler. Thus, it does not have the limitation of the prior work. Fig. \ref{fig:4} compares the transmission response of the MI that has been measured against the Verilog-A based simulation. The simulated frequency spectrum is obtained by passing the wavelength as a parameter to each component and running wavelength sweep simulation. Also, in order to focus on the actual device response, we remove the grating coupler effects with the help of de-embedding structures. Following this, we adjust the waveguide effective index and loss to match the measurements. One simulation was run using the bidirectional signaling methodology introduced above whereas the other was executed without this feature. It is clear that neglecting the bidirectional signal propagation leads to erroneous results.

\begin{figure}[h]
\centering\includegraphics[width=11cm]{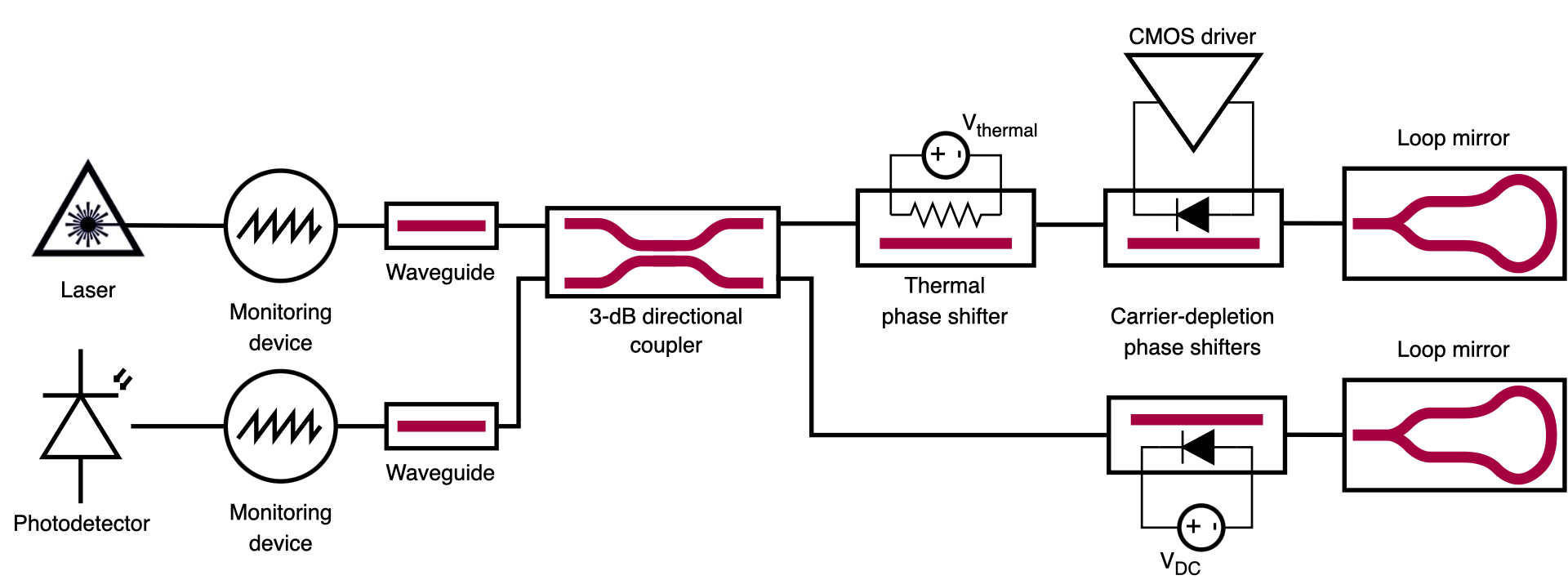}
\caption{Schematic diagram of the MIM built using Verilog-A components.}
\label{fig:3}
\end{figure}

\subsection{Michelson Interferometer Modulator testbench}

\begin{figure}[t!]
\centering\includegraphics[width=7cm]{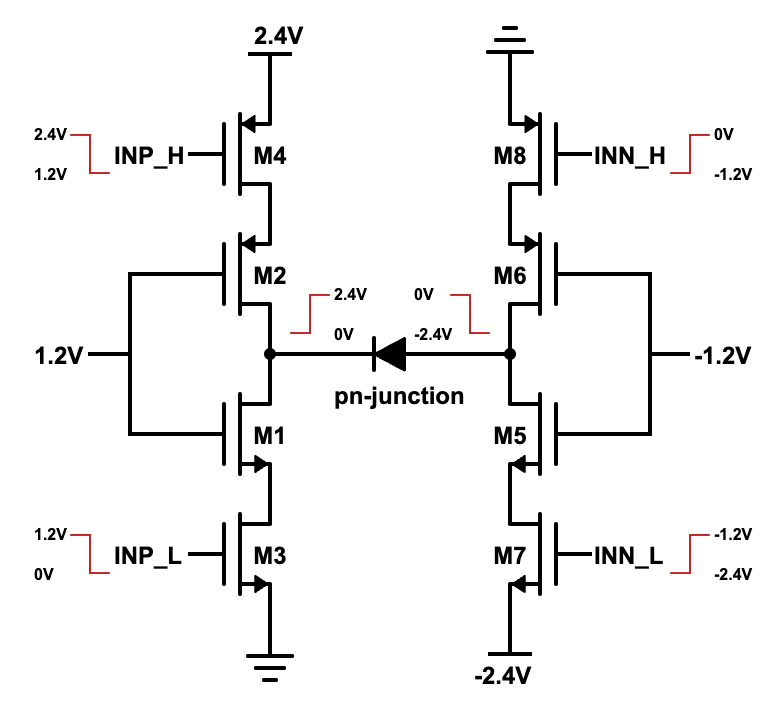}
\caption{Pseudo-differential inverters based CMOS driver circuit.}
\label{fig:3b}
\end{figure}

\begin{figure}[t!]
\centering\includegraphics[width=13cm]{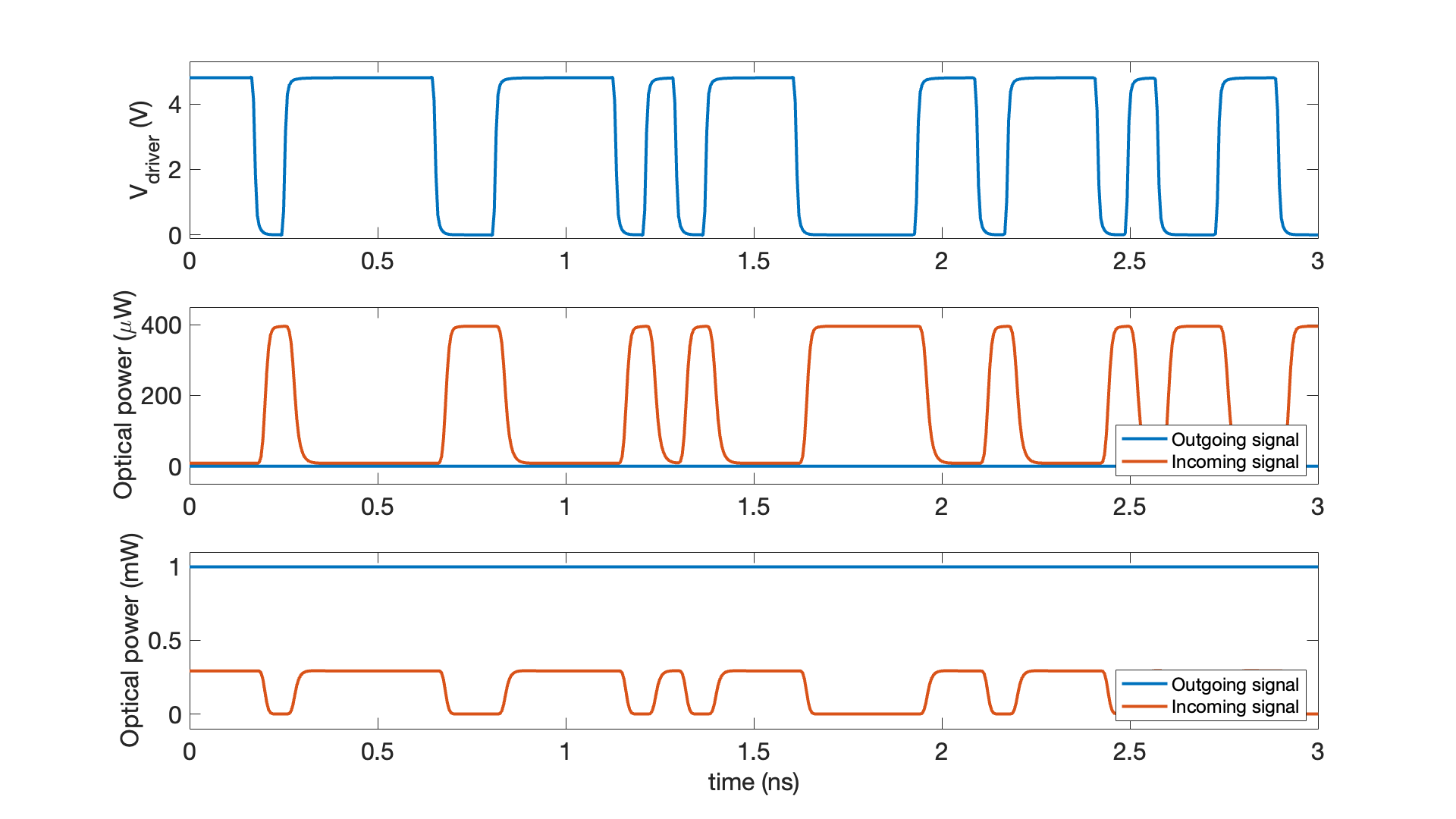}
\caption{Transient analysis results of the MIM. The top plot shows the differential output voltage of the CMOS driver. The middle and bottom plots illustrate the incoming and outgoing signals at the photodetector and laser, respectively.}
\label{fig:5}
\end{figure}

For transient analysis, we replace the waveguides in each arm of the previous example with the thermal and carrier-depletion phase shifters. We also add a transistor-level electronic driving circuit from the 65 nm CMOS technology library to modulate the carrier-depletion phase shifter. Fig. \ref{fig:3} shows the schematic view of the Michelson Interferometer Modulator (MIM) simulated in the same environment as the CMOS driver. The detailed schematic of the CMOS driver circuit is displayed in Fig. \ref{fig:3b}. It is based on pseudo-differential inverters topology \cite{haoli}, which can provide 4.8V differential output voltage swing. The voltage swing requirements are dictated by the pn-junction phase shifter length, which is 800 $\mu m$ in this design. M1, M3, M5, and M7 are n-type MOSFETs, whereas M2, M4, M6, and M8 are p-type MOSFETs. The input signal arrangement is also shown in the same figure. The basic idea here is to alternately switch on a set of transistors (either M1, M3, M6, M8 or M2, M4, M5, M7) to create 0V-4.8V modulation voltage across the pn-junction diode, which, in its turn, changes the effective index of the doped waveguide, thereby modulating the amount of phase shift applied to the optical signal. The transient behavior of the circuit is provided in Fig. \ref{fig:5}. The reflected signals arrive at the laser and photodetector interfaces in accordance with the modulation signal. Furthermore, as expected, the laser's output signal (i.e. forward propagating signal) remains the same regardless of the incoming signals. Similarly, the photodetector's forward propagating signal is zero at all times.

\begin{figure}[h]
\centering\includegraphics[width=10cm]{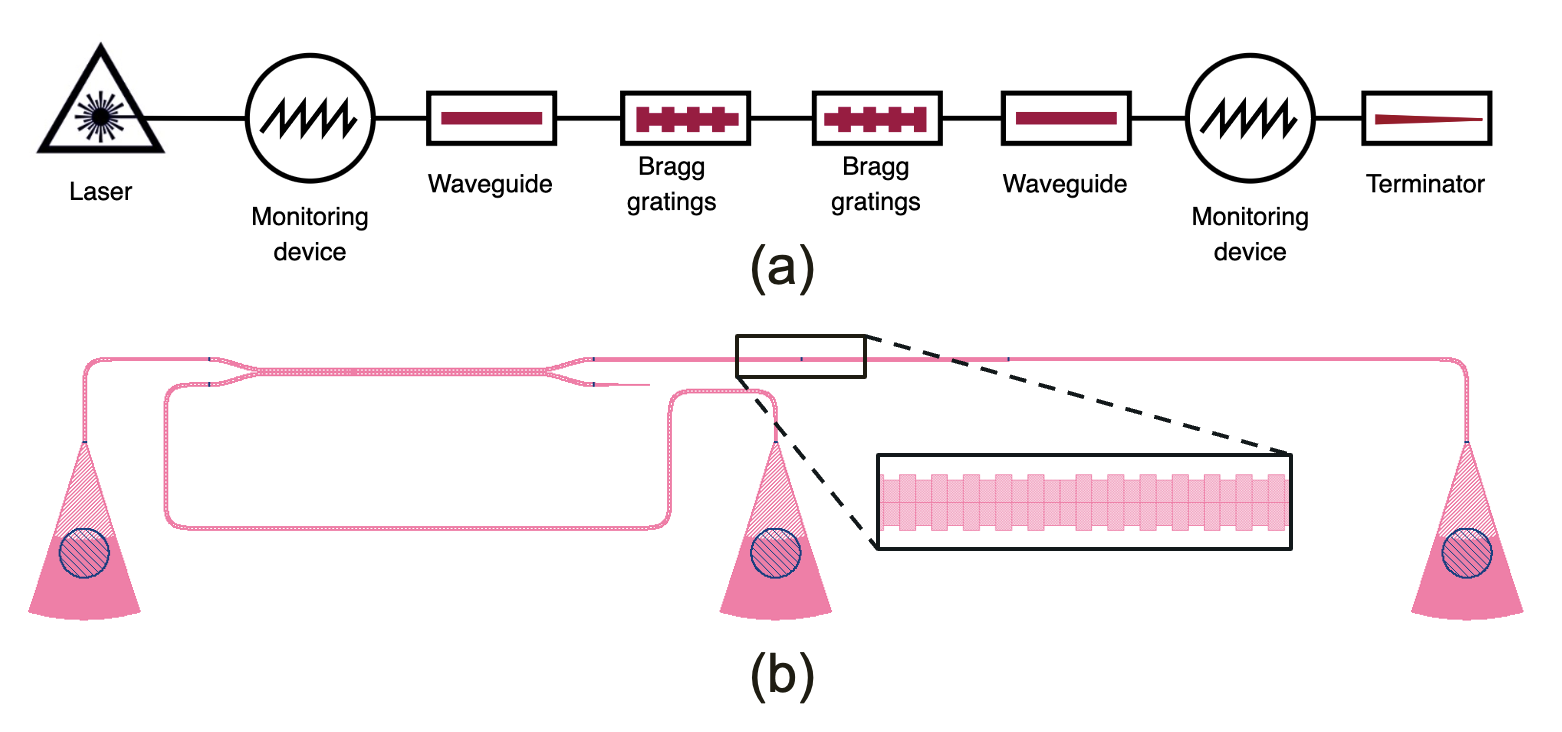}
\caption{(a) Schematic diagram of the Bragg gratings based Fabry-Perot cavity; (b) Layout of the measured Fabry-Perot cavity.}
\label{fig:6}
\end{figure}

\begin{figure}[h]
\centering\includegraphics[width=13cm]{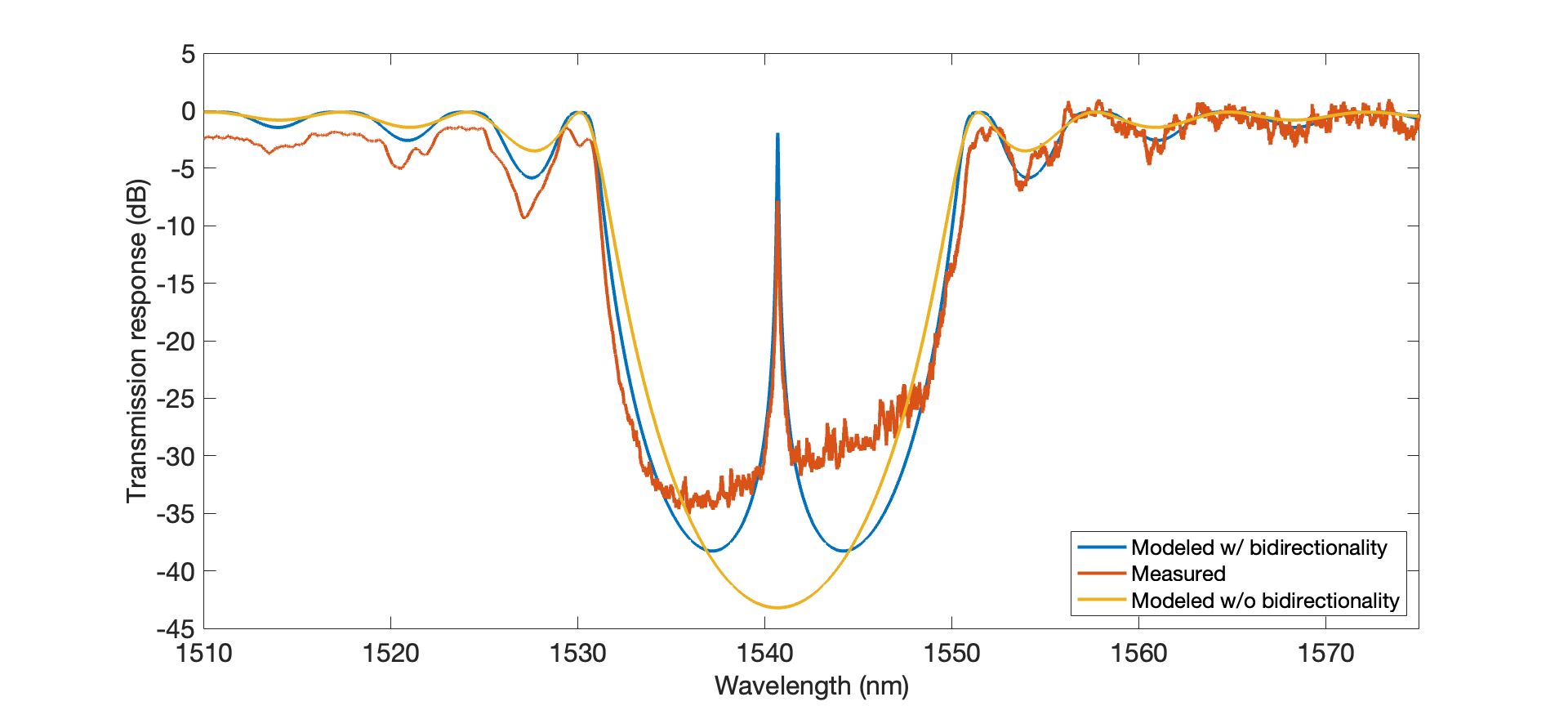}
\caption{Bandpass response of the Fabry-Perot cavity.}
\label{fig:7}
\end{figure}

\begin{figure}[h!]
\centering\includegraphics[width=13cm]{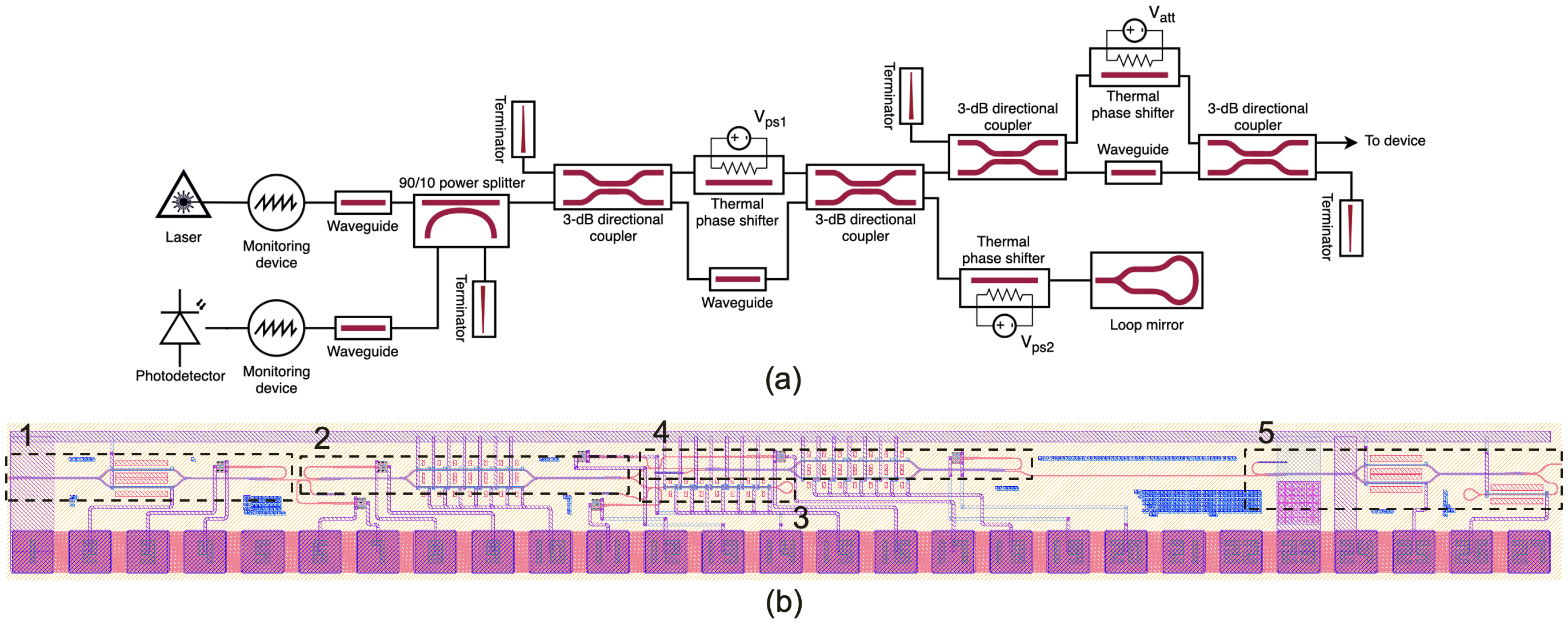}
\caption{(a) Schematic diagram of the reflection cancellation circuit (RCC); (b) Layout of the measured RCC (presented with author's permission \cite{Shoman_2020}). In part (b): 1 - routing adjustable splitter connecting laser output to the RCC; 2 - intermediate adjustable splitter; 3 - phase tunable reflector; 4 - output stage adjustable splitter; 5 - control device generating the undesired reflections.}
\label{fig:8}
\end{figure}

\begin{figure}[h!]
\centering\includegraphics[width=12cm]{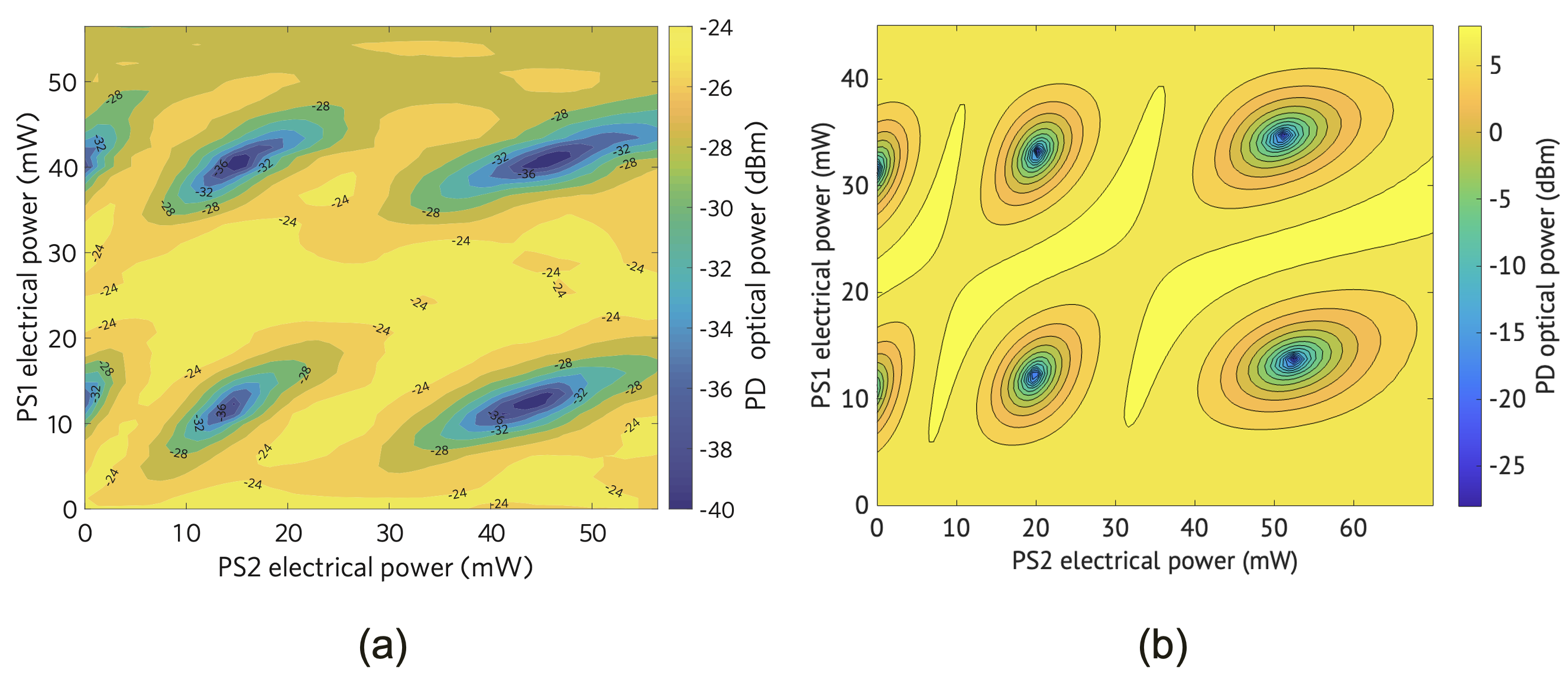}
\caption{(a) Measured 2D sweep of the first two phase shifters (PS1 and PS2); (b) Simulated 2D sweep of the same phase shifters. The measured response is reprinted from \cite{Shoman_2020} with author's permission. }
\label{fig:9}
\end{figure}

\subsection{Fabry-Perot Cavity testbench}

Next, we consider a testbench for the Bragg gratings based Fabry-Perot Cavity (FPC) simulation, similar to what is presented in \cite{Patel_Ghillino_Korthorst_2021}, wherein the Verilog-A photonic models were deemed incapable of simulating bidirectional signal propagation. Fig.~\ref{fig:6} illustrates the developed testbench (Fig.~\ref{fig:6}a) and the layout of the corresponding fabricated device (Fig.~\ref{fig:6}b), where we designed the FPC to have a resonant peak at the wavelength close to 1541 nm. The fabricated device has 120 corrugations per reflector, each with a width of 50 nm, whereas the cavity length is equal to one Bragg period (317 nm). Since FPC is a circuit consisting of passive elements only, its baseband time domain response is static. Therefore the main focus is on its frequency spectrum. We run wavelength sweep simulation on the presented circuit and compare it against the measured FPC response to arrive at Fig.~\ref{fig:7}. The difference between the bidirectionality enabled simulation and conventional simulation is apparent. The latter fails to capture the bandpass behaviour at around 1541 nm, which may be critical for some applications. Similar to the previous testbench, the grating coupler effect is de-embedded from the actual device response and the Verilog-A models here are also fitted to the measurement results. In this way, the proposed Verilog-A modeling fully captures the true FPC response, in contrast to the limitation described in~\cite{Patel_Ghillino_Korthorst_2021}. 

\subsection{Reflection Cancellation Circuit testbench}

The final simulation scenario involves the creation of the testbench for photonic Reflection Cancellation Circuit (RCC), illustrated in Fig. \ref{fig:8}a. If a fraction of the optical power emitted from the laser reflects back into the laser, it can lead to the broadening of its linewidth, increased Relative Intensity Noise (RIN), and eventually render the laser unstable. An RCC is a compact on-chip solution that ensures the stability of the laser by actively canceling out any reflections to the laser through its phase tunable reflector arm and the interference effect~\cite{ShomanJLT}. More details of this circuit can be found in~\cite{ShomanJLT}. It consists of: 1) 90/10 power tap connecting to the photodetector to track the reflection cancellation amount; 2) intermediate adjustable splitter ($V_{ps1}$) along with phase tunable reflector ($V_{ps2}$) to control the magnitude and phase of the reflection canceling signal; and 3) output stage adjustable splitter ($V_{att}$) used to control the amount of reflection coming back from the subsequent device. Fig. \ref{fig:8}b demonstrates the layout, which, besides the RCC itself, contains routing and reflection generator stages (highlighted as 1 and 5 in the figure). While the output portion of the RCC is only used for the initial setup, the former two are critical for an effective cancellation of any reflection coming from the device. We use the $V_{ps1}$ and $V_{ps2}$ knobs to run 2D sweep of the reflection landscape and observe the optical signal strength at the input of the photodetector. Fig. \ref{fig:9} shows the resulting measurement and bidirectional signaling-enabled simulation response. The simulation results closely follow the real RCC behaviour, including the skewed response which comes from the imbalanced 3-dB directional couplers, and the asymmetric power requirements to reach the global minima which is the evidence of thermal crosstalk between the phase shifters \cite{Shoman_2020}. The thermal crosstalk is modeled by introducing a crosstalk parameter into the voltage source of each thermal phase shifter. Note that in the simulations we do not include grating couplers and therefore achieve higher optical power readings. Additionally, one can observe deviations in the electrical power requirements necessary to arrive at a global minima between the two plots. This could be attributed to the fact that our thermal phase shifter model is purely analytical and may have different thermal characteristics. We emphasize that without the bidirectional signaling it would not be possible to capture local or global minima, which is essential for RCC's functionality.

\section{Conclusion}

We have presented an approach to model photonic devices in Verilog-A. It handles forward and backward propagating signals in a single wire, removing the redundancy of the prior arts. Unlike prior art~\cite{DEFOUCAULD2023108538}, it also supports multiple optical sources. Furthermore, our method enables integration of simulation and measurement data into Verilog-A models through s-parameters. In this way, one can capture response of any arbitrary passive photonic components. The simulation results showed the efficacy of our solution thereby making electro-optic co-simulation more convenient and accurate.

\begin{backmatter}

\bmsection{Funding} 
This work was supported in part by the Natural Sciences and Engineering Research Council of Canada and in part by Schmidt Sciences.

\bmsection{Acknowledgements} 
The authors would like to thank CMC Microsystems and Roozbeh Mehrabadi for access to CAD tools and technology, Hossam Shoman for access to the RCC measurement data and layout, and the staff from the online course “Phot1x: Silicon Photonics Design, Fabrication and Data Analysis” offered via edX UBCx for the chip fabrication and measurements.

\bmsection{Disclosures} The authors declare no conflicts of interest.
\end{backmatter}



\bibliography{Optica-template}






\end{document}